\documentclass[%
 reprint,
 amsmath,amssymb,
 aps,
 prl,
]{revtex4-1}

\usepackage{graphicx}
\usepackage{dcolumn}
\usepackage{bm}
\begin{document}

\title{Rapid and accurate molecular deprotonation energies from quantum alchemy}

\author{Guido Falk von Rudorff}
\affiliation{Institute of Physical Chemistry and National Center for Computational Design and Discovery of Novel Materials (MARVEL), Department of Chemistry, University of Basel, Klingelbergstrasse 80, CH-4056 Basel, Switzerland}

\author{O. Anatole von Lilienfeld}
\email{anatole.vonlilienfeld@unibas.ch}
\affiliation{Institute of Physical Chemistry and National Center for Computational Design and Discovery of Novel Materials (MARVEL), Department of Chemistry, University of Basel, Klingelbergstrasse 80, CH-4056 Basel, Switzerland}

\date{\today}

\begin{abstract}
We assess the applicability of Alchemical Perturbation Density Functional Theory (APDFT) for quickly and accurately estimating deprotonation energies. We have considered all possible single and double deprotonations in one hundred small organic molecules drawn at random from QM9 [Ramakrishnan et al, JCTC 2015]. Numerical evidence is presented for 5'160 deprotonated species at both HF/def2-TZVP and CCSD/6-31G* level of theory. 
We show that the perturbation expansion formalism of APDFT quickly converges to reliable results: using CCSD electron densities and derivatives, regular Hartree-Fock is outperformed at second or third order for ranking all possible doubly or singly deprotonated molecules, respectively. 
CCSD single deprotonation energies are reproduced within 1.4\,kcal/mol on average within third order APDFT. 
We introduce a hybrid approach were the computational cost of APDFT is reduced even further by mixing first order terms at a higher level of theory (CCSD) with higher order terms at a lower level of theory only (HF). We find that this approach reaches 2 kcal/mol accuracy in absolute deprotonation energies compared to CCSD at 2\% of the computational cost of third order APDFT.
\end{abstract}

\maketitle

\section{Introduction}

\begin{figure}
    \centering
    \includegraphics{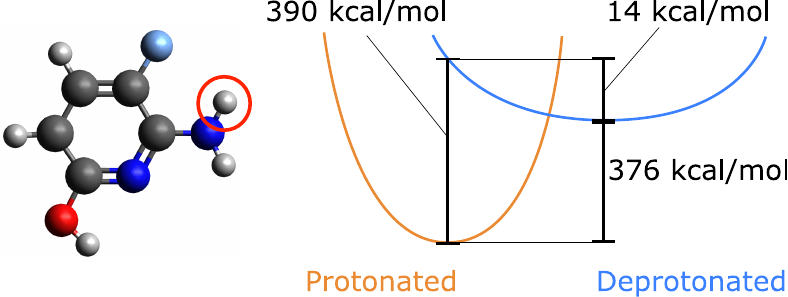}
    \caption{Schematic potential energy surfaces for a protonated and singly deprotonated molecule (left, deprotonated site indicated). The vertical and relaxed deprotonation energies are shown. Data calculated at HF/6-31G*.}
    \label{fig:schema}
\end{figure}

The proton affinity as an inherent property of a molecule with determines its protonation state, determines the enthalpic contribution to the pKa\cite{Moser2010,Sulpizi2010}, determines reaction dynamics\cite{Carlin2016}, and impacts proton transport\cite{Carlin2016,Rudorff2016}. Evaluating which sites have the lowest energetic barrier for deprotonation is one part of predicting the overall protonation state of a molecule. The proton affinity $E_\text{pa}$ is given as 
\begin{align}
E_\text{pa} &\equiv -\Delta H = \Delta E + \Delta E_\text{ZPVE} + H(\text{H}^+)\\
    \Delta H &= H(\text{AH}) - H(\text{A}^-) - H(\text{H}^+)
\end{align}
where $H$ is the enthalpy ($\frac{5}{2}RT$ for the free proton), $\Delta E$ is the dominating contribution of the total energy change in deprotonation, and $\Delta E_\text{ZPVE}$ is the zero-point vibrational energy contribution. It is commonly assumed that the difference in zero-point vibrational energy between the neutral molecule and the anion is small\cite{Klamt2003}, even though there is numerical evidence of this being far from a general rule\cite{Range2005}.  However, the zero-point vibrational energy and configurational energy differences as shown in Figure~\ref{fig:schema} can nowadays be modeled quite accurately with conventional universal force-fields or semi-empirical methods, or even with quantum machine learning (See Ref.~\cite{Faber2018} for an example). For this study, we focus on the dominating  total energy contribution, which is most susceptible to the local electronic structure and therefore requires accurate quantum chemistry methods. 

Previous work has shown that only high levels of theory afford deprotonation energies which are accurate enough to allow comparison to experiment with chemical accuracy\cite{Range2006}. These calculations however are expensive, since almost all practically relevant molecules can be deprotonated at multiple sites which drastically increases the computational cost. For example, in the case of the QM9 database\cite{Ramakrishnan2014,Ruddigkeit2012} which contains organic molecules with up to nine heavy atoms (not counting hydrogens), on average nine protons are available per molecule. 
If up to two sites are allowed to be deprotonated, this yields $9+9\cdot 8/2 = 45$ possible protonation states. 
For larger molecules where the protonation state is relevant e.g. for molecular packing\cite{Zhang2015,Sadhukhan2015} or conformational structure of proteins\cite{Russo2012} this number quickly becomes so large that the systematic enumeration of all protonation states is rendered computationally unfeasible.

\begin{figure*}
    \centering
    \includegraphics[width=\textwidth]{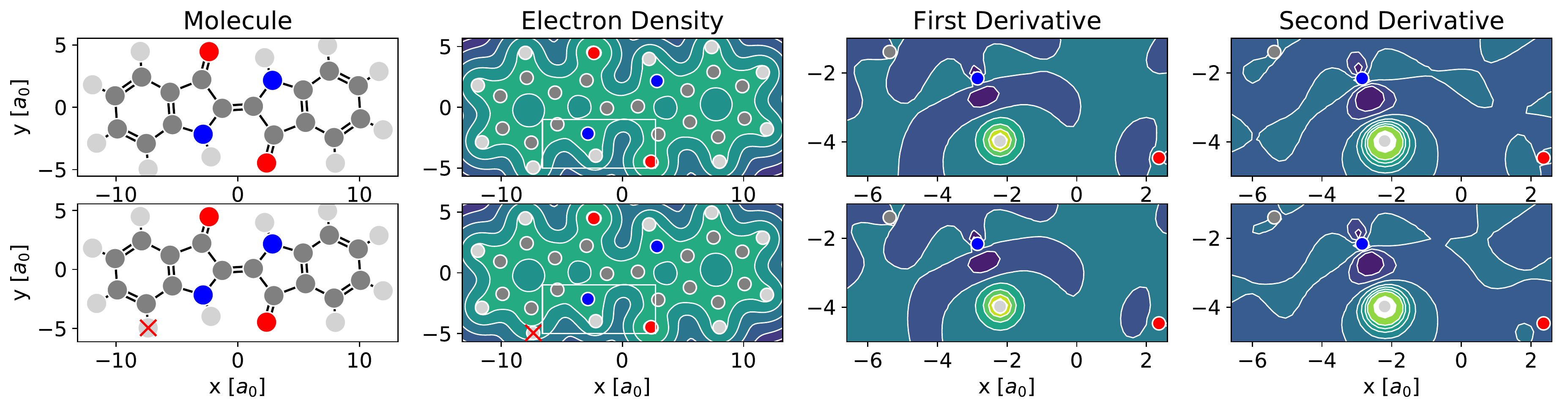}
    \caption{Illustration of locality of alchemical single deprotonation derivatives in indigo molecule (top row) and indigo anion deprotonated at one site (bottom row, site marked with red cross). Molecular structure, and contour plot of slices of electron density $\rho$ and its first and second alchemical deprotonation electron density derivatives $\partial_\lambda\rho, \partial^2_\lambda\rho$. Zoom-in  only shown in the non-negligible domain around the deprotonation site, marked by the white rectangle in the electron density plot. Positive derivative values shown in yellow, negative in blue. All data obtained at HF/6-31G level.}
    \label{fig:derivatives}
\end{figure*}

Recently, Alchemical Perturbation Density Functional Theory (APDFT)\cite{Rudorff2018} has been developed which offers a way to drastically reduce computational cost of such screening efforts. 
The core idea is to treat a change in nuclear charges as a perturbation to a molecular Hamiltonian where all other degrees of freedom such as geometry and numbers of electrons are fixed. 
This is achieved by defining a new mixed molecular Hamiltonian $\hat H$
\begin{align}
    \hat H \equiv \lambda \hat  H_t + (1-\lambda)\hat H_r
\end{align}
consisting of a linear interpolation between the molecular Hamiltonians of reference and target molecule, respectively. The interpolation is driven by a coupling parameter $\lambda$, similar to the adiabatic connection picture. See also Refs.~\cite{Foldy1951,Wilson1962, Epstein1967, Politzer1974,Politzer1987,Politzer2002,anatole-prl2005, ArianaAlchemy2006, anatole-jcp2006-2, anatole-jcp2009-2, LesiukHigherOrderAlchemy2012, anatole-ijqc2013, Samuel-CHIMIA2014, munoz2017predictive, Fias2018} for the background of quantum based computational alchemy. 
While less commonly used, these methods have by now already been demonstrated to reach useful accuracy in many cases. 
Specific examples include estimated 
changes in HOMO eigenvalues of benzene due to BN doping~\cite{anatole-jcp2007},
hydration free energies of ions~\cite{anatole-jcp2009},
adsorption of small molecules on metal clusters~\cite{CatalystSheppard2010},
energies of mixed metal clusters~\cite{WeigendSchrodtAhlrichs2004, weigend2014extending},
energies of mixed ionic crystals~\cite{AlchemyAlisa_2016}, 
transition metal solid properties~\cite{MoritzBaben-JCP2016},
covalent binding in single, double, and triple bonds of small molecules\cite{Samuel-JCP2016},
small molecule adsorption to catalytic surfaces~\cite{saravanan2017alchemical, griego2018benchmarking},
water adsorption on BN doped graphitic materials~\cite{Yasmine-JCP2017}, 
electronic locality within molecules~\cite{StijnPNAS2017},
BN doping in C$_{60}$~\cite{Balawender2018},
band-gap engineering in GaAlAs semi-conductors~\cite{Samuel2018bandgaps},
energies in BN substituted benzene and coronene derivatives, as well as all III-V and IV-IV solids based on perturbations of Ge~\cite{Fias2018}.

Contrary to the typical computational quantum alchemy application which modifies nuclear charges or pseudo-potentials of heavy elements, we here focus on the annihilation of protons only. 
More specifically, the overall difference between the energy $E_t$ of a target molecule, i.e.~any of the many possible deprotonated anions, and the energy $E_r$ of the neutral reference molecule can be written according to~\cite{Rudorff2018} as

\begin{align}
E_\text{t} - E_\text{r}  & = \Delta E_\text{NN} + \int d\mathbf{r} \Delta v\sum _{n=1}^{\infty }{\frac {1}{n!} \left.\frac{\partial^{n-1} \rho_\lambda}{\partial \lambda^{n-1}}\right\rvert_{\lambda=0}} \;  \label{eqn:apdft}
\end{align}

where $\Delta E_\text{NN} = \sum_I Z_I/|{\bf R}_{\rm H}-{\bf R}_I|$ is the nuclear repulsion of the annihilated proton, i.e. just the difference in nuclear-nuclear interaction between reference and target molecule; 
$\Delta v$ is the change in the nuclear Coulomb potential going from reference to target molecule, and $\partial_\lambda\rho$ gives the density derivatives in the direction of the interpolation path described by $\lambda$. Note that the density derivatives are evaluated at the reference molecule (i.e. $\lambda =0$) only. In recent work, we have shown\cite{Rudorff2018} that this infinite sum converges rather quickly, meaning that the first few terms recover the vast majority of the energy change between reference and target molecule and even allow decomposition into atomic energy as well as electron density contributions~\cite{Rudorff2019a}. In practice, this means that its sufficient to evaluate the electron density and its first few derivatives for the base molecule only, so there is no combinatorial scaling with the total number of protonation sites in this method.

So far, the density derivatives $\partial_\lambda\rho$ implicitly depend on the target compound, since they denote the density derivatives in the direction of the target molecule. However, by virtue of the chain rule, we can express the first two orders as
\begin{align}
\frac{\partial\rho}{\partial\lambda} &= \sum_I\frac{\partial\rho}{\partial Z_I}\frac{\partial Z_I}{\partial\lambda} = \sum_I\frac{\partial\rho}{\partial Z_I}\Delta Z_I\\
\frac{\partial^2\rho}{\partial\lambda^2} &= \sum_J\sum_I\frac{\partial^2\rho}{\partial Z_I\partial Z_J}\frac{\partial Z_I}{\partial\lambda}\frac{\partial Z_J}{\partial\lambda} \nonumber\\
&= \sum_J\sum_I\frac{\partial^2\rho}{\partial Z_I\partial Z_J}\Delta Z_I\Delta Z_J
\end{align}
where $I$ and $J$ run over all nuclei, $Z_I$ denotes the charge of nucleus $I$, and $\Delta Z_I$ is the corresponding difference between reference and target molecule on site $I$. 

In the context of APDFT, deprotonation is equivalent to changing the nuclear charge of the Hydrogen site to zero while keeping the total number of electrons fixed. This means that either $\Delta Z_I=0$ (for heavy atoms or protons that stay in place for a given target) or $\Delta Z_I=-1$ (for sites which are deprotonated).

\section{Methods}

\begin{figure}
    \centering
    \includegraphics[width=\columnwidth,clip, trim=8cm 11cm 5cm 0cm]{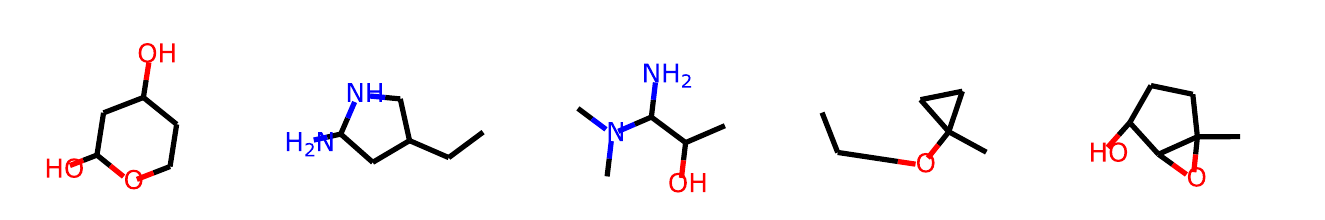}
    \includegraphics[width=\columnwidth]{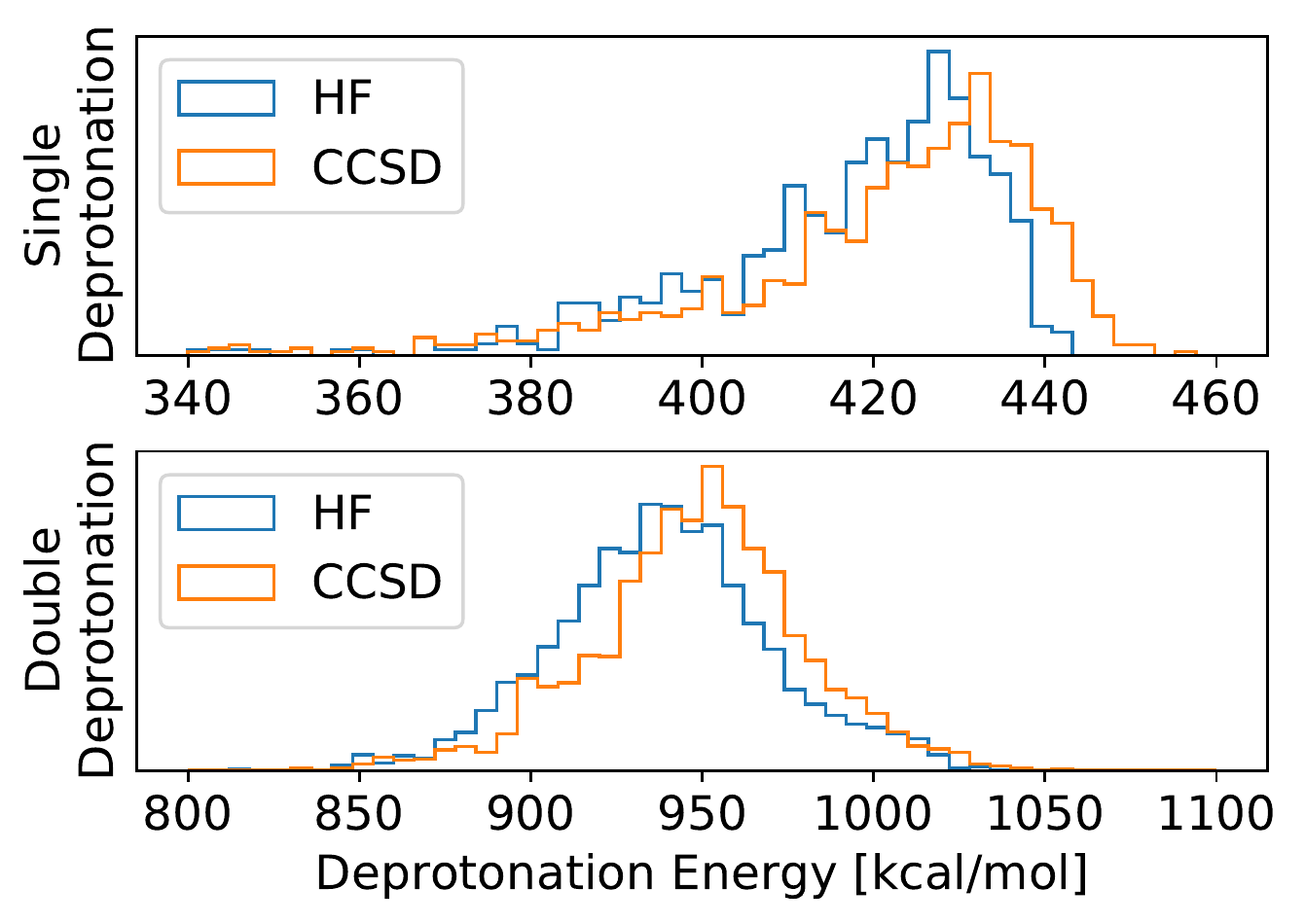}
    \caption{Top: Example molecules used for the evaluation of HF, CCSD, and APDFT (full list in the SI). Bottom: Histogram of the deprotonation energies for single and double deprotonation in the data set considered. Data shown for HF/def2-TZVP and CCSD/6-31G*.}
    \label{fig:mollist}
\end{figure}

To numerically assess this approach, we chose 100 random molecules (full list in the SI, five examples in Figure~\ref{fig:mollist}) from the QM9 database\cite{Ramakrishnan2014} in the B3LYP local minimum geometries given in that database. All of these molecules have been evaluated on two levels of theory: HF/def2-TZVP\cite{Weigend2005} and CCSD/6-31G*\cite{Ditchfield1971,Hariharan1973,Hehre1972} as provided by the BasisSetExchange\cite{Pritchard2019,Feller1996,Schuchardt2007}. For CCSD, we chose a smaller basis set to reduce the overall computational cost of the self-consistent results to which we compare our APDFT results. The def2-TZVP basis set is parametrically optimized at the HF level, i.e. partial derivatives of the basis set parameters are designed to be zero at HF level. This makes the def2 family particularly accurate for APDFT at HF level, which is why it has been chosen in this work.

We used the APDFT code\footnote{https://github.com/ferchault/APDFT} and PySCF\cite{PYSCF} to calculate the electron density and its first two derivatives for both levels of theory (details in the SI). For a subset thereof, the electron densities and their derivatives have been validated with both MRCC\cite{mrcc} and Gaussian\cite{Gaussian09} and the same corresponding levels of theory.

For each molecule, all unique singly and doubly deprotonated configurations have been evaluated explicitly (i.e. self-consistently) by iterating over all sites, in total 5'160 for each level of theory. All these evaluations have been done vertically, i.e. in the geometry of the fully protonated molecule as found in the QM9 database. Upon deprotonation, the basis functions of the hydrogen atoms in question have been removed together with the nucleus. The density derivatives are obtained from central finite differences where the nuclear charges are perturbed by $0.05e$, which requires 1'816 calculations for all molecules for the first order and 8'304 calculations for the second order contributions. Note that none of these molecules feature intramolecular hydrogen bonds (IMHB) in the geometries we investigated. With the energy contribution of IMHB being significant for relative ranking of conformers but much smaller in magnitude than the overall deprotonation energy, we expect some but no large differences between the alchemical derivatives for removing a proton that is and one that is not part of an IMHB.

All integrals have been evaluated analytically by calculating the electrostatic potential at the nuclei for all obtained electron densities.

\section{Results and Discussion}

As per eqn.~\ref{eqn:apdft}, APDFT requires the electron density derivatives w.r.t. the nuclear charges. Figure~\ref{fig:derivatives} shows how these derivatives look like for hydrogen sites. One can think of these derivatives being the electron density response upon adding a proton at that location. In the first derivative, electron density gets concentrated around the proton, which is to satisfy Kato's cusp theorem\cite{Kato1957} that any nuclear charge needs to create a singularity in the electron density. The electron density that is built up around the Hydrogen atom mostly comes from the atom it is bonded to and (to a lesser extent) from the bond axis. The second derivative (which has a smaller absolute magnitude than the first derivative) then polarises the electron density around the Hydrogen atom more strongly by depleting the electron density at the side facing the bonded atom and accumulating density at the opposite side. This means that the vast majority of the density rearrangement upon protonation is happening along the bond axis of that Hydrogen and, as such, is highly local. Interestingly, this applies to both first and second order density derivatives in a part of the molecule that is in close vicinity to regions of high electron density like the Oxygen atom. Since the change in energy obtained via APDFT depends on these density derivatives only, the observation of highly localised electron density derivatives is an indication of deprotonation energies being additive. As shown in Figure~\ref{fig:derivatives}, the density derivatives due to a deprotonation are largely unaffected by already deprotonated sites nearby. This means that the electron density derivatives constituting the second step in formation of a doubly deprotonated molecule are still highly localised and similar to a single deprotonation event. This points towards a high transferability of the density derivatives across molecular environments.

\begin{figure}
    \centering
    \includegraphics[width=\columnwidth]{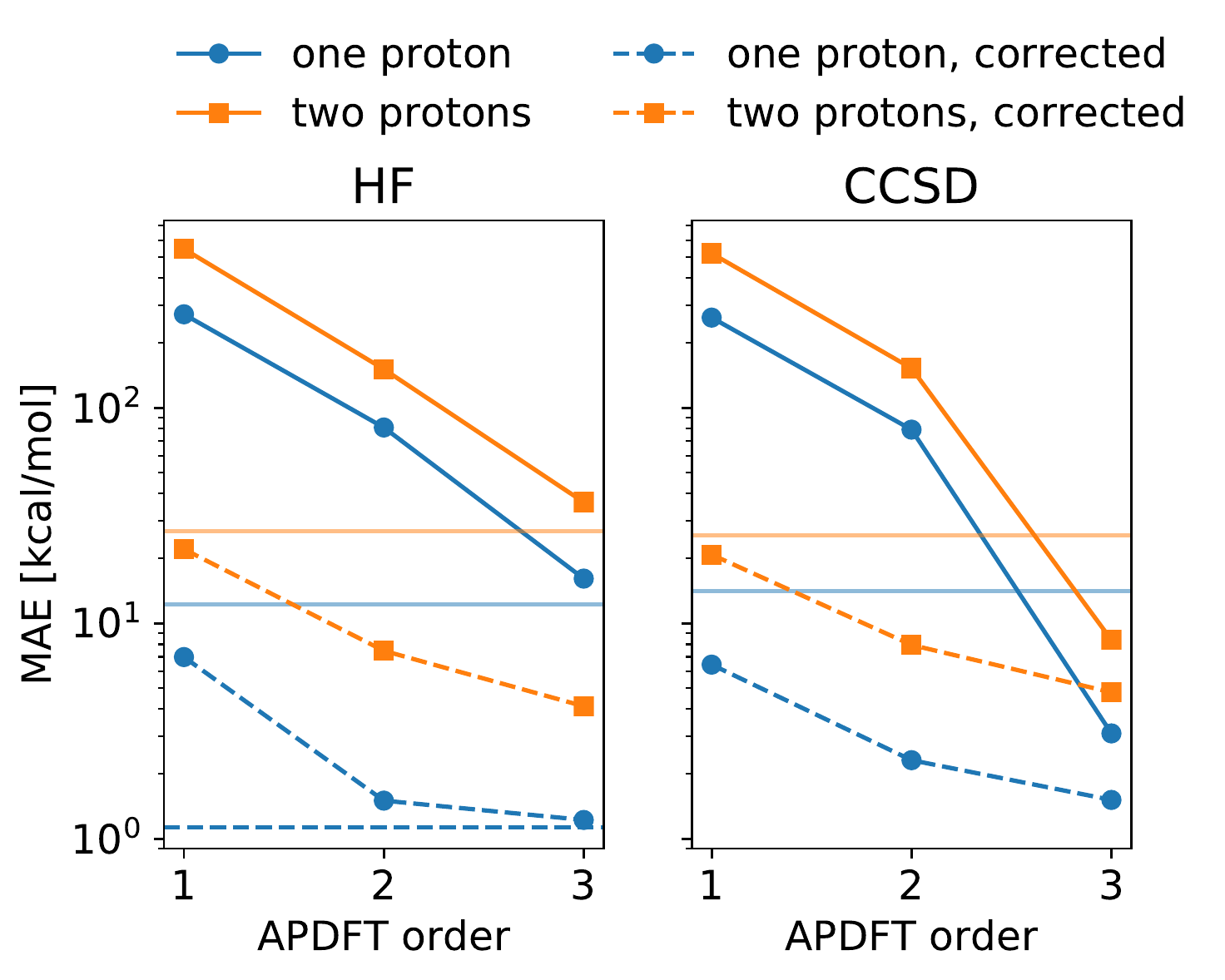}
    \caption{Mean absolute error (MAE) for deprotonation energies obtained via APDFT with expansion order when compared to the self-consistent energies at the same level of theory. Data set split for single deprotonation and double deprotonation. Stroked horizontal lines show the standard deviation of the deprotonation energies for comparison. The dashed horizontal line is the HF error compared to CCSD results. Correction for median error discussed in the text. Left panel shows data for HF/def2-TZVP, right panel shows CCSD/6-31G*. }
    \label{fig:order}
\end{figure}

The energy expression of APDFT, eqn.~\ref{eqn:apdft}, is a sum from infinitely many terms. With more and more higher order terms, the expression becomes more accurate, but also more expensive to evaluate. To be practically relevant, this sum needs to be quickly converging. Figure~\ref{fig:order} shows the mean absolute error (MAE) for APDFT systematically decreasing with the order $n$ in eqn.~\ref{eqn:apdft}, since more and more of the electron density response is taken into account. This is the case for both singly and doubly deprotonated molecules which are separated in the Figure. Consistently, i.e. regardless of method and APDFT order, stripping two protons from the molecule carries a significantly larger error. This is because if two sites are alchemically changed at the same time, these changes interact. In APDFT, the first two orders only contain per-site terms while the third order is the first to contain pairwise terms. 

The residual error of APDFT however is systematic in nature. This allows for a simple correction where each value is shifted by the median error of comparable results, which captures the average contributions from higher orders in the APDFT expression. This correction brings down the MAE by one order of magnitude. With 1.4 kcal/mol accuracy, APDFT is quantitative for the deprotonation energy of one site comparable to HF which reaches a residual error of 1.3 kcal/mol. In practice, when searching for site-specific deprotonation energies, this only requires a few calibrating calculations to find the median error for a given APDFT order which then can applied to the remaining data set. Since the median is a robust metric, i.e. only marginally affected by outliers, very few such calibration calculations will stabilise the value for this correction.

To set this accuracy into perspective, Figure~\ref{fig:mollist} shows the histogram of deprotonation energies for the molecules in our data set. They span 120\,kcal/mol for single deprotonation and 200\,kcal/mol for double deprotonation. By comparison, the APDFT accuracy of 1.4\,kcal/mol is nearly two orders of magnitude smaller than the value range.


In the context of APDFT, the success of this simple correction can be understood physically. Let us consider the case where the first two orders are included explicitly for single deprotonation. Then the energy expression is 

\begin{align}
\Delta E  =& \Delta E_\text{NN} + \int d\mathbf{r}\Delta v  \left[\rho + \frac{1}{2}\left.\frac{\partial\rho}{\partial \lambda}\right\rvert_{\lambda=0}\right] + \nonumber\\
&\sum _{n=2}^{\infty }\frac {1}{n!} \int d\mathbf{r} \Delta v{\left.\frac{\partial^{n-1} \rho_\lambda}{\partial \lambda^{n-1}}\right\rvert_{\lambda=0}}
\end{align}
where the last sum contains all higher order terms. If their mean is constant regardless of target as seen by the success of the correction, then the integral over change in external potential $\Delta v$ and density derivative must have a strong contribution that does not depend on the actual target molecule. Since $\Delta v$ is always a $1/r$ function centered on the proton in question, the only variable component is the density derivative. For all targets, it has the same relative position w.r.t. $\Delta v$. Therefore the fact that the integrals for all higher orders are largely constant regardless of the molecule in question means that the spatial shape of the electron density derivatives is mostly identical for all protonation sites. As soon as multiple sites are deprotonated via an alchemical transformation, the change in external potential $\Delta v$ includes the change in the second site and thus is no longer exactly the same for different target molecules. Therefore the correction should be less effective for double deprotonation--which indeed is the case as shown in Figure~\ref{fig:order}. While the aforementioned mathematical argument only holds for the mean value, in practice one would prefer the median as a robust estimator of the mean, since only few observations will be used to obtain the mean error.

To set the MAE into perspective, Figure~\ref{fig:order} also shows the standard deviation of the deprotonation energies in our dataset. If one were to estimate deprotonation energies by their average, an error of that magnitude would be expected. Interestingly, the absolute values without the correction only come close to (HF) or improve upon (CCSD) this level of accuracy at third order APDFT. After the correction however, even first order APDFT is better than this estimate even though first order APDFT carries nearly no computational cost and only uses the electron density of the reference molecule. While the first order term is not sufficient to obtain practically useful deprotonation energies, this illustrates how quickly relevant physics is captured in the sum of the APDFT energy expression.

\begin{figure}
    \centering
    \includegraphics[width=\columnwidth]{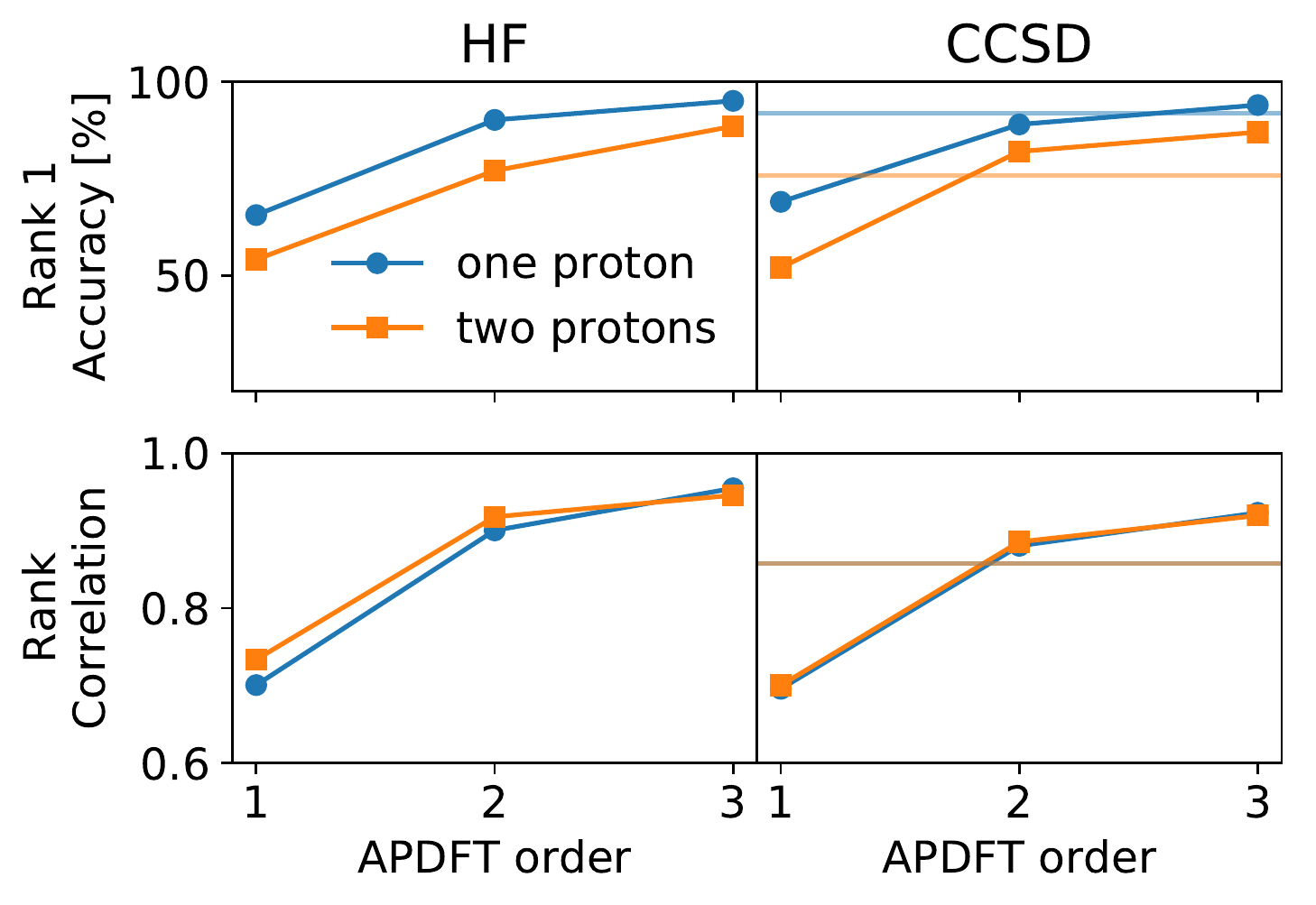}
    \caption{Performance of APDFT in ranking deprotonation sites within a molecule. Top row: rank 1 accuracy, meaning the percentage of cases where the most stable proton site has been correctly identified. Bottom row: Kendall's $\tau$ rank correlation coefficient describing the overall ranking accuracy. Perfect agreement is reached for a value of 1, perfect anti-correlation for -1. First column for HF/def2-TZVP, second column for CCSD/6-31G*. Horizontal lines in the second column denote the performance of HF for the same metric. All predictions corrected by median residual error as discussed in the text.}
    \label{fig:rank}
\end{figure}

This correction however is only required if absolute deprotonation energies are required. The typical use case is to identify the one proton of a molecule that can be stripped away most easily. This requires ranking the deprotonation energies of all sites in a given molecule. Figure~\ref{fig:rank} shows the performance of APDFT in this regard. Interestingly, the rank 1 accuracy exceeds 50\,\%\ even at first order, i.e. without the inclusion of any density derivative at all. For both levels of theory investigated in this work, the accuracy reaches 94\,\%\ after inclusion of the median-corrected third order of APDFT for single deprotonation. This is remarkable, since HF itself is correct in 85\,\%\ of the cases when CCSD ranking is the reference. This way, using APDFT on CCSD data is more accurate than doing all calculations self-consistently with Hartree-Fock. Therefore, it can be more efficient to invest into few higher-quality calculations and then use APDFT for the derivatives for all individual targets than to brute-force the enumeration over all possible targets at some intermediate level of theory.

For two protons being removed at the same time, the APDFT predictions systematically improve with order as well. Note that the ranking improvement by including the third order terms is not as large as the improvement seen in Figure~\ref{fig:order} for absolute energies. This points towards the first two orders recovering the overall ranking and the third order mostly shifting deprotonation energies to be more accurate.

Figure~\ref{fig:rank} also shows Kendall's $\tau$ as metric of the overall accuracy of the ranking, not only of one particular rank. Kendall's $\tau$\cite{Kendall1938} has been chosen since it is more resilient against effects of small numbers of ranks to consider than e.g. the Spearman rank. The picture for the overall ranking is very consistent with the rank 1 accuracy: starting from second order terms, APDFT on CCSD data is more accurate than self-consistent Hartree-Fock calculations when compared to the CCSD reference results. Again, the ranking improvement of the third order terms is noticeable but a sufficient ranking accuracy is already established at second order.

\begin{figure}
    \centering
    \includegraphics[width=\columnwidth]{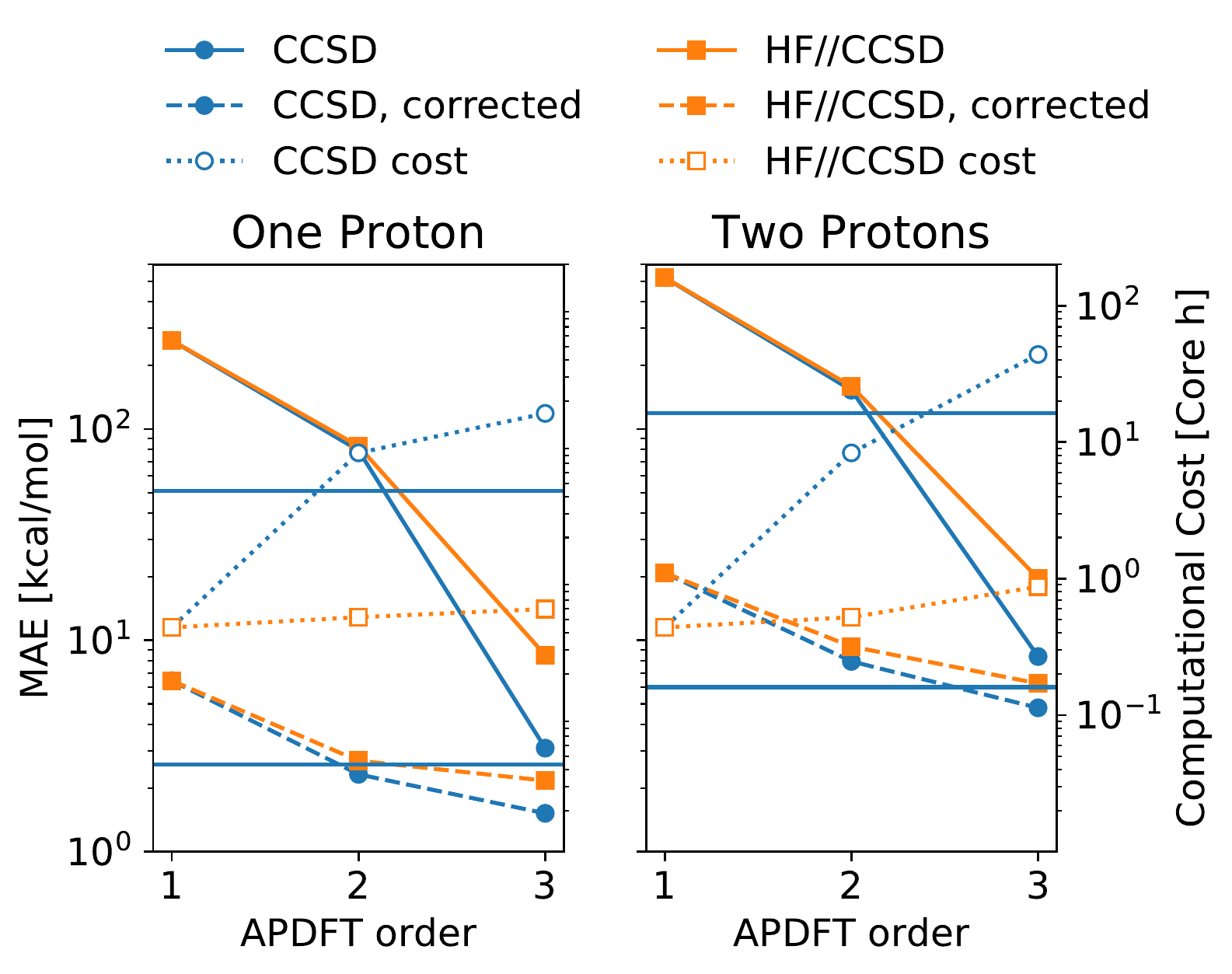}
    \caption{Mean absolute error (MAE) of deprotonation energies obtained from APDFT with CCSD/6-31G* data for first order and HF/def2-TZVP data for higher orders, denoted HF//CCSD. Exact expression given in eqn.~\ref{eqn:mixed}. Median-corrected (see text) data shown as dashed lines. Computational cost of APDFT shown with dotted lines. Upper/lower horizontal line refer to brute-force full SCF with CCSD/6-31G* and HF/6-31G*, respectively. Reference data is CCSD/6-31G* deprotonation energies for both panels.}
    \label{fig:mixed}
\end{figure}

Generally, the spatial electron densities are quite similar across methods, while the total energies associated with these densities vary widely. If the electron densities are similar between methods, then their derivatives must be similar as well in order to keep that similarity across chemical space. In the APDFT energy expression, the first order term establishes the total energy baseline for any target molecule while the higher order terms give density-based corrections to that first estimate. Note that no total energy of any level of theory enters the expression for the higher order terms. Now if densities and their derivatives are more similar between methods than the total energies are, then it is a promising route to obtain the first order term from high quality calculations (e.g. CCSD) and the higher orders being approximated by the density derivatives obtained at a lower and cheaper level of theory (e.g. HF). 

To this end, we shall give any density or density derivative with the level of theory at which it has been obtained as superscript. Then the first three orders for the deprotonation energy $\Delta E$ as obtained from central finite difference derivatives with a finite difference stencil of $\Delta \lambda$ are given by
\begin{align}
\Delta E  \approx& \Delta E_\text{NN} + \int d\mathbf{r}\Delta v  \left.\left[\rho^\text{CCSD} + \frac{1}{2}\frac{\partial\rho^\text{HF}}{\partial \lambda}+\frac{1}{6}\frac{\partial^2\rho^\text{HF}}{\partial \lambda^2}\right]\right\rvert_{\lambda=0} \nonumber\\
=& \Delta E_\text{NN} + \int d\mathbf{r}\Delta v \left[\rho^\text{CCSD}+\frac{\rho^\text{HF}(\Delta\lambda)-\rho^\text{HF}(-\Delta\lambda)}{4\Delta\lambda}\right.\nonumber\\
&\left.+\frac{\rho^\text{HF}(\Delta\lambda)-2\rho^\text{HF}+\rho^\text{HF}(-\Delta\lambda)}{6\Delta\lambda^2}\right]\label{eqn:mixed}
\end{align}
Note that the electron density of the neutral molecule is required at both levels of theory in order to obtain consistent higher-order derivatives.

Figure~\ref{fig:mixed} shows the resulting accuracy for deprotonation energies following this approach. In direct comparison to CCSD density derivatives, this mixed approach is of comparable quantitative accuracy. Moreover, the median correction outlined above is still applicable for the results of mixed levels of theory. Since the difference between CCSD and HF is the inclusion of correlation energy in the former, this means that the correlation energy needs to be highly similar between different deprotonated targets, even though it can vary arbitrarily between neutral molecules. Despite the purely Coulombic expression in eqn~\ref{eqn:apdft}, APDFT recovers all energy contributions covered by the level of theory at which the density derivatives have been evaluated. Consequently, the electron density derivatives only include physical effects that are part of the level of theory at which they have been evaluated. Therefore this mixed approach is only likely to work for those cases where the correlation energy is substantial enough to require the inclusion of it in the first order but also locally constant in chemical space, i.e. of comparable value for nearby target molecules.

As shown in Figure~\ref{fig:mixed}, this mixed approach requires 2\%\ of the computational cost of third order APDFT for the 6-31G* basis set used. Note that this speedup becomes more and more pronounced with larger basis sets and larger molecules, since the inherent scaling of CCSD is worse than the scaling of HF w.r.t. the number of basis functions. While for very small molecules, a brute-force calculation of all possible single deprotonations can be cheaper than APDFT, since the number of derivatives APDFT requires is comparably high in small molecules, the hybrid approach HF//CCSD is always significantly cheaper. Most importantly, due to the chain rule trick, APDFT scales with the combinatorial increase of possible deprotonations for multiple protons being removed: Figure~\ref{fig:mixed} shows second order non- APDFT to be more expensive than brute-force CCSD for \textit{single} deprotonations, but already for double protonations, APDFT is cheaper. The hybrid approach however, is computationally more efficient in all cases.

\section{Conclusion}
In the context of deprotonation of small organic molecules, this work suggests the use of the quantum alchemy method APDFT to quantify deprotonation energies $\Delta E$ and rank the individual sites by using high-quality reference calculations and the density derivatives only instead of calculating deprotonated species explicitly with a medium level method. 
If required, the computational cost can be reduced further by evaluating higher order derivatives at a lower level of theory. The systematic contribution of higher order terms in APDFT that are not evaluated at all can be treated by shifting results by their molecule-independent median deviation from reference results. In the case of CCSD and HF, this procedure yields more reliable results at a substantially lower computational cost.

The accuracy for absolute deprotonation energies of 1.4\,kcal/mol are on par with quantum chemical calculations with a large basis sets when compared to experiment\cite{Moser2010} and substantially outperform semiempirical methods\cite{Range2005}. In terms of ranking, the quantum alchemy predictions from APDFT based on CCSD derivatives are found to be more accurate than explicit HF calculations. This means that APDFT gives energies close to the explicity calculated reference values which in turn are closer to experiment if the level of theory is able to capture more relevant physical effects. This could be particularly helpful for cases like metal centers where only high-level reference methods are able to describe the electronic structure sufficiently accurately.

As an outlook, our findings are also promising for enabling ensemble calculations of free energies throughout chemical compound space, generating extensive lists of pK$_a$ estimates~\cite{pKaLore2008} for entire molecular  libraries. Future work will deal with more systematic assessments of the hybrid approach for larger sets of molecules.

\begin{acknowledgments}
We acknowledge support by the Swiss National Science foundation (No.~PP00P2\_138932, 407540\_167186 NFP 75 Big Data, 200021\_175747, NCCR MARVEL).
Some calculations were performed at sciCORE (http://scicore.unibas.ch/) scientific computing core facility at University of Basel.
\end{acknowledgments}

\bibliography{main}
\end{document}